\documentstyle[12pt,epsfig]{article}

\begin{document}


     
\begin{centering}
\bigskip
{\Large \bf A phenomenological description of quantum-gravity-induced
space-time noise}\\
\bigskip
\bigskip
\bigskip
{\bf Giovanni AMELINO-CAMELIA}\\
\bigskip
Dipartimento di Fisica, Universit\'{a} ``La Sapienza", P.le Moro 2,
I-00185 Roma, Italy\\ 
\end{centering}
\vspace{1cm}
     
\baselineskip = 24pt

\noindent

\vspace{1cm} 
\begin{center} 
{\bf ABSTRACT} 
\end{center} 
 
{\leftskip=0.6in \rightskip=0.6in  
I propose a phenomenological description
of space-time foam and discuss the experimental limits
that are within reach of forthcoming experiments.} 
  
\newpage 
 
\baselineskip 12pt plus .5pt minus .5pt 

\pagenumbering{arabic} 
\pagestyle{plain}  

``Space-time foam" is a geometric picture
of the smallest size scales of the Universe, which is
characterized mainly by the presence of quantum
uncertainties in the measurement of distances.
All quantum-gravity theories should have
some kind of foam~\cite{wheely,hawkfoam}, but 
the description of foam varies according to 
the theory. Experimental observations 
establishing some of the properties of space-time foam
would provide a crucial hint for the search of the 
correct quantum gravity.
I previously showed~\cite{gacgwi} that 
foam-induced distance fluctuations
would affect gravity-wave
interferometers by introducing a new source of noise, 
but the present level of development of candidate theories
of quantum gravity does not allow~\cite{bignapap} to derive 
detailed distance-fluctuation
predictions to guide the work of experimentalists.
Here I propose a phenomenological approach that describes
directly space-time foam and this new approach 
naturally leads to a picture of quantum distance fluctuations 
that is independent of the specific setup of a given interferometer.
The only unknown in the model is the
length scale that sets the overall magnitude of the effect. 
I find that recent data~\cite{ligoprototype,tama}
already rule out the possibility that this length
scale be identified with the ``string length''
($10^{-34} m < L_s < 10^{-33}m$).
Experiments that will soon start operating will probe
values of the length scale even smaller than the ``Planck length'' 
($L_p \sim 10^{-35}m$).

One of the most robust~\cite{wheely,hawkfoam}
expectations for quantum gravity,
as the theory describing the interplay
between gravity and quantum mechanics,
is that space-time itself at the smallest scales 
should manifest quantum fluctuations of geometry,
which could be described in terms of
a highly non-trivial 
structure of (3+1-dimensional) space-time.
This can be roughly visualized using an analogy
with ordinary ``foamy" or ``spongy" materials,
imagining however that physical processes are confined to
the material of the ``sponge".
Another useful intuition-building analogy can be made
with the Heisenberg uncertainty principle: while that
principle assigns a minimum on uncertainties relevant
for the combined measurement of
the position and the momentum of a particle
in a fixed (background) space-time, we now expect an uncertainty
principle for space-time itself. This would
set an absolute limit on the measurability of distances.

It is natural~\cite{gacgwi,bignapap,ahlunature,nggwi,polonpap}
to characterize operatively this space-time foam
through its implications for an ideal interferometer.
Quantum fluctuations of distances would
be observed in an interferometer as a source of noise.
Theoretical predictions for this noise
could be tested by comparing them with the
noise levels actually found experimentally.
In particular, a given picture of foam-induced distance fluctuations
is of course ruled out if it predicts more noise than
the total noise seen experimentally.

The first studies on this 
subject~\cite{gacgwi,bignapap,ahlunature,nggwi,polonpap}
indicated in various ways that the sensitivity of modern interferometers
could be sufficient for the detection of space-time fluctuations
originating at Planckian distance scales. Of course, in order 
to  provide guidance to the experimentalists involved in
interferometric tests, it would be useful to have a detailed 
description of the fluctuations induced by space-time foam.
Unfortunately, the scarcity of experimental information on the
quantum-gravity realm has not yet allowed a 
proper ``selection process", so there are a large number of
quantum-gravity candidates. Moreover, even the two approaches whose
mathematical/logical consistency
has been already explored in some depth,
the one based on ``critical superstrings''~\cite{string1,string2}
and the one based on ``canonical/loop quantum
gravity''~\cite{cqgab,cqgcar,cqglee},
have not yet matured a satisfactory understanding
of their physical implications, 
such as the properties of space-time foam. 
In the few phenomenological programmes investigating
other quantum properties of 
space-time~\cite{ehns,elmn1,elmn2,elmn3,grbgac}
the difficulties deriving from the preliminary status 
of quantum-gravity theories have been 
circumvented by developing
direct phenomenological descriptions of the relevant phenomena.
I propose to apply the same strategy to the
description of the noise induced in interferometers
by quantum gravity.

My task is partly facilitated by the fact that
in order to guide interferometric studies
of foam it is only necessary to estimate
a relatively simple (single-variable) function:
the power spectrum $\rho_h(f)$ 
of the strain noise~\cite{saulson,rwold}.
[Strain here has the standard
engineering definition $h \equiv \Delta L/L$ 
in terms of the displacement $\Delta L$ in a given
distance $L$.] In fact, the strain noise power spectrum,
through its dependence on the frequency $f$ at which 
observations are performed,
contains the most significant information
on the distance fluctuations, such as the mean square deviation
(which is given by the integral of the power spectrum over
the bandwidth of operation of the detector),
and is the quantity against which the observations are compared.

The quantum-gravity-induced strain noise should depend only
on the Planck length, 
the speed-of-light constant $c$ ($c \simeq 3 {\cdot} 10^8 m/s$),
and, perhaps, a length scale characterizing the properties
of the apparatus with respect to quantum gravity.
I observe that within this conceptual framework
there is a unique compellingly-simple candidate
for a foam-induced ``white noise''
(noise with constant, $f$-independent, power spectrum).
White noise is to be expected whenever
the relevant stochastic phenomena
are such that there is no correlation between one
fluctuation and the next, an hypothesis which appears rather
plausible for the case of space-time fluctuations.
The hypothesis that foam-induced noise be white is also
consistent with the the intuition emerging from
analogies~\cite{garaythermal} between thermal environments
and the environment provided by foam as a (dynamical)
arena for physical processes.
According to these studies one can see foam-induced noise as
essentially analogous to thermal noise in various physical
contexts (such as electric circuits, where noise
is generated by the thermal agitation of the electrons), 
which is indeed white whenever the bandwidth of interest is below 
some characteristic (resonant) frequency.
In the case of foam-induced noise the characteristic
frequency (which should be somewhere in the neighborhood of
the quantum-gravity frequency scale $c/L_p$)
would be much higher than the frequencies of operation
of our interferometers, and foam noise would be white
at those frequencies.

Within a white-noise model, by observing that the
strain noise power spectrum carries dimensions of $H\!z^{-1}$,
one is naturally led to the estimate 
\begin{equation}
 \rho_h(f) = {\rm constant} \sim  {L_p \over c} 
\sim 5 {\cdot} 10^{-44} H\!z^{-1} ~.
\label{white}
\end{equation}

I also observe that, since, as mentioned,
the frequencies we can access experimentally
are much smaller than $c/L_p$, 
white noise is actually the only admissable
structure for foam-induced strain noise within the hypothesis that
this noise be independent of the characteristics of the apparatus
which is used as a space-time probe.
In fact this hypothesis implies that $\rho_h$
can only depend on its argument $f$,
on the Planck length
and on the speed-of-light constant,
and therefore the most general low-frequency expansion
is of the type
\begin{equation}
 \rho_h(f) =
a_0 {L_p \over c} + a_1 \left( {L_p \over c} \right)^2 f
+ a_2 \left( {L_p \over c} \right)^3 f^2 + ...
\label{expans}
\end{equation}
where the $a_i$ are numerical coefficients
and all monomials of the type $f^{-|n|}$
were not included in the expansion because they would require
coefficients of the type $L_p^{-|n|+1}$ (which would be inconsistent
with the fact that quantum-gravity effects must disappear in
the limit  $L_p \rightarrow 0$).
For $f \ll c/L_p$
the expansion (\ref{expans}) is well approximated by its first term, 
which corresponds to the dimensional estimate (\ref{white}).
From the point of view of experimental tests
it is also important to consider the value of the coefficient $a_0$,
{\it i.e.} to take into account the inherent uncertainty
associated with the dimensional estimate (\ref{white}).
In this type of studies
based on dimensional analysis,
the natural guess, which often turns out to be correct, 
is that $a_0$ is of order 1, but it is not uncommon to find
a disagreement between the dimensional estimate and the experimental
result of a few orders of magnitude.
In testing (\ref{white}) we shall therefore be looking for sensitivities
extending a few orders of magnitude below the $L_p/c$ level.

In the same sense that the estimate
(\ref{white}) provides a compelling candidate for
foam-induced noise in quantum-gravity theories with
ordinary point-like (particle)
fundamental objects, in theories with extended ({\it e.g.}
string-like) fundamental objects characterized
by a length scale $L_s$ it appears natural to consider
the low-frequency estimate
\begin{equation}
\rho_h \sim  {L_s \over c} ~.
\label{whitestring}
\end{equation}
In string theories $L_s$ would be the string length,
which is expected to be
somewhere between a factor 10 and a factor 100 larger
than the Planck length, and therefore for $L_s/c$ there is a
range of 
values $5 {\cdot} 10^{-43} H\!z^{-1} < L_s/c < 5 {\cdot} 10^{-42} H\!z^{-1}$.

Since they predict no dependence on the nature of the apparatus being
used to probe space-time, these estimates (\ref{white}) and
(\ref{whitestring})
can be tested using any detector with sensitivity to distance strain,
such as interferometers and resonant-bar detectors.
Remarkably,
in spite of the smallness of the effects predicted,
these types of experiments are reaching such a high level
of sensitivity that (\ref{white}) and (\ref{whitestring})
are going to be completely tested (either discovered
or ruled out) within a few years.

Denoting with $\rho_h^{TOT}$ the total
strain noise power spectrum observed by the experiments,
the present level of interferometric data
is best characterized
by the results obtained  
by the {\it 40-meter interferometer}~\cite{ligoprototype}
at Caltech and the {\it TAMA interferometer}~\cite{tama} at the
Mitaka campus of the Japanese National Astronomical Observatory,
both reaching $\rho_h^{TOT}$ of order $10^{-40} H\!z^{-1}$
(the lowest level has been achieved by {\it TAMA} around 
1$kH\!z$: $\rho_h^{TOT} \sim 3 {\cdot} 10^{-41} H\!z^{-1}$).
Even more remarkable is the present sensitivity
$\rho_h^{TOT} \simeq 5 {\cdot} 10^{-43}H\!z^{-1}$ 
of resonant-bar
detectors such as NAUTILUS~\cite{nautilus}
(which achieved it near $924 H\! z$).
This is already quite close to 
the natural quantum-gravity
estimate $L_p/c$ of (\ref{white}),
and is already at the level $L_s/c$.
We are already probing a potentially interesting region and in
order to complete a satisfactory test of the estimates
(\ref{white}) and (\ref{whitestring}) we only need
to improve the sensitivity by a few orders of magnitude
(in order to exclude also the possibility that 
the coefficient $a_0$ be somewhat smaller than 1).

This will be accomplished in the near future.
Planned upgrades of the NAUTILUS resonant-bar detector
are expected~\cite{nautilus,micgwb}
to reach sensitivity
at the level $7 {\cdot} 10^{-45} H\!z^{-1}$.
The LIGO/VIRGO generation of 
interferometers~\cite{ligo,virgo} 
should achieve sensitivity of the 
order of $10^{-44} H\!z^{-1}$ within a year or two,
during its first phase of operation.
A few years later, with
the space interferometer LISA~\cite{lisa}
and especially with the ``advanced 
phase''~\cite{micgwb,ligo} of the LIGO/VIRGO interferometers,
another significant sensitivity improvement should be achieved:
according to recent estimates~\cite{ligo} it should be
possible to reach sensitivity levels in the neighborhood
of $10^{-48} H\!z^{-1}$, more than four orders of
magnitude below the natural $L_p/c$ estimate here
considered! 

This expected experimental progress is described in the figure
together with the $L_p/c$ white-noise level and the
analogous noise-level predictions that can be obtained
by assuming instead that the foam-induced noise be
of ``random-walk" type ({\it i.e.} with $f^{-2}$ frequency
dependence of the power spectrum~\cite{rwold}).
Through the example of random-walk noise the figure shows
that the sensitivity of modern interferometers is significant
also with respect to non-white models 
of foam-induced noise.
This is an important consideration in assessing
the overall significance of the interferometric studies here
considered; in fact,
the quantum-gravity realm is very far from the experimental
contexts that formed our intuition,
and, while the simple $L_p$-linear white-noise model may appear 
natural at present,
it is reassuring that this experimental programme can explore
a rather wide class of noise models.

The example of random-walk noise
can also be used to illustrate what
would be the implications of having noise that,
unlike $L_p$-linear white noise,
necessarily depends on some
experiment-characteristic
length scale $\Lambda$.
A model with
random-walk strain noise 
linearly suppressed by the Planck
length would have to predict a power spectrum of the
form $\rho_h \sim c L_p f^{-2} \Lambda^{-2}$.
Our capability to test such a model
is to be described with the range of values of $\Lambda$
which we can exclude.
As shown in the figure,
for the $L_p$-linear random-walk-noise model
the excluded range of values of $\Lambda$ extends all the way
up to values of $\Lambda$ of the order of the optical length
of the arms of the interferometer.
In the random-walk
case we will soon even reach some sensitivity to
models with effects quadratically suppressed by the
Planck length; in fact,
as shown in the figure, the LISA
interferometer~\cite{lisa} will be able to test the possibility of
noise levels of the
type $\rho_h \sim c L_p^2 f^{-2} \Lambda^{-3}$
for plausible values of the experiment-characteristic
length scale $\Lambda$.
Since other quantum-gravity-motivated experimental programmes
can only achieve sensitivity to effects linear in the
Planck length~\cite{polonpap,ehns,elmn1,elmn2,elmn3,grbgac},
LISA's capability to reach ``$L_p^2$ sensitivity"
will mark the beginning of another significant phase
in the search of quantum properties of space-time.

In order to render more powerful the phenomenological approach
here proposed the most urgent challenge to theory concerns the
understanding of the role of energy considerations
in quantum gravity. 
If one applied a similar phenomenological approach to the analysis
of other interferometric noise sources, it would be easy to
find ways to discriminate between different noise models 
by evaluating the amount of energy required by
the corresponding fluctuation schemes~\cite{micgwb}.
Unfortunately, while energy considerations are rather elementary
when the analysis is supported by a fixed background space-time,
the fact that quantum gravity cannot rely on a background space-time 
renders energy considerations much more subtle~\cite{thooftnew}.
The role that this issue could play in the development
of the approach here proposed
will be discussed in detail elsewhere~\cite{longnat}.

On the experiment side, my analysis of quantum-gravity noise
provides additional motivation for the studies planned
by LISA and by the ``advanced phase"
of the LIGO/VIRGO interferometers. The original classical-gravity
objective of modern interferometers, the discovery
of Einstein's gravity waves,
might well be achieved already by the ``first phase"
of LIGO/VIRGO, but
even in that case it appears to be necessary to maintain 
the present ambitious sensitivity objectives for LISA and 
the LIGO/VIRGO ``advanced phase". The payoff could be
the first experimental evidence of a quantum property 
of space-time.

\baselineskip 24pt plus .5pt minus .5pt

\newpage

\begin{figure}[p] 
\begin{center} 
\epsfig{file=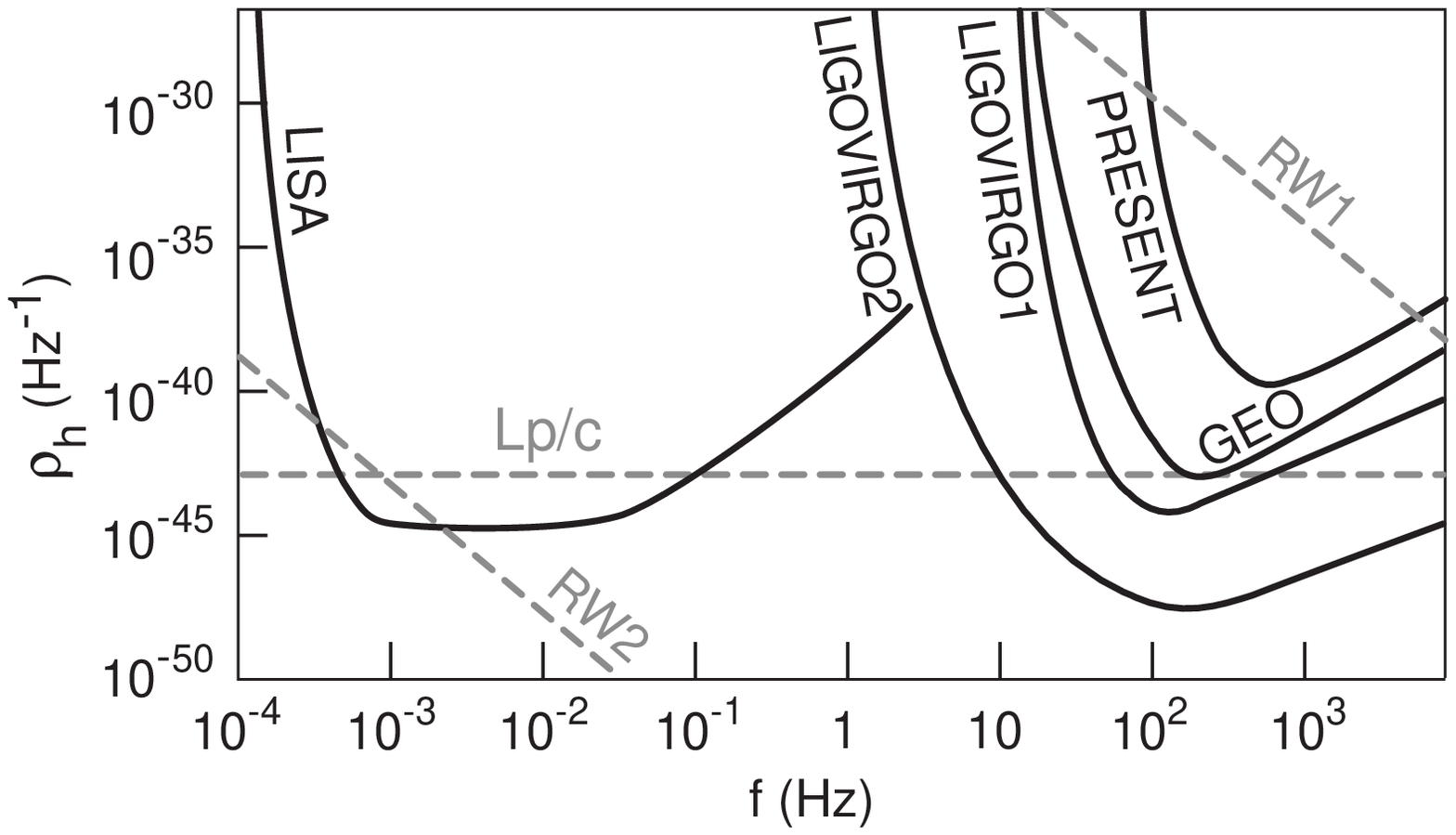,height=12truecm}\quad 
\end{center} 
\caption{A qualitative
(at best semi-quantitative) comparison
between the sensitivity of certain interferometers 
and the types of
strain noise power spectra here considered.
The evolution from the level of sensitivity (``PRESENT'')
of interferometers already in operation,
to GEO (``GEO'') and
the first phase of the LIGO and VIRGO interferometers (``LIGOVIRGO1''),
and finally to LISA (``LISA'')
and the second phase of LIGO and VIRGO (``LIGOVIRGO2''), 
will take us through some significant phenomenological milestones
among candidate foam-induced noise levels.
The white-noise line ``$L_p/c$'' will be crossed already by 
the first phase of LIGO and VIRGO.
The line ``RW1'' is representative of the random-walk scenario 
with magnitude suppressed linearly by the Planck length,
and, as mentioned, is ruled out by ``PRESENT'' data.
The figure also shows that with LISA 
we will even start probing a small
range of values of the overall coefficient $c/\Lambda^3$
(where $\Lambda$
should be a scale characteristic of the experimental setup)
of the scenario with random-walk noise levels suppressed 
by the square of the Planck  length. In fact,
the line ``RW2'' corresponds 
to $\rho_h \sim c L_p^2/(\lambda^3 f^2)$,
where $\lambda$ is the wavelength of the LISA beam.}
\label{fig1} 
\end{figure}

\vfil


\begin{thebibliography}{99}
     
\bibitem{wheely} Wheeler, J.A.~in {\it Relativity, groups and topology},
(eds.~De Witt, B.S. \& De Witt, C.M.) 
(Gordon and Breach, New York, 1963).

\bibitem{hawkfoam} Hawking, S.W.~{\it Spacetime foam},
Nuc.~Phys.~B144, 349-362 (1978).

\bibitem{gacgwi} Amelino-Camelia, G.~{\it Gravity-wave 
interferometers as quantum-gravity detectors},
Nature 398, 216-218 (1999).

\bibitem{bignapap} Amelino-Camelia, G.~{\it Gravity-wave
interferometers as probes of a
low-energy effective quantum gravity},
Phys.~Rev.~D62, 024015 (2000).

\bibitem{ligoprototype} Abramovici, A.~{\it et al}, {\it Improved 
sensitivity in a gravitational wave interferometer and implications 
for LIGO}, Phys.~Lett.~{A218}, 157-163 (1996).

\bibitem{tama} Updated information
on the progress of observations performed by the
TAMA interferometer can be found
at the WWW site http://tamago.mtk.nao.ac.jp/.

\bibitem{ahlunature} Ahluwalia, D.V.~{\it Quantum gravity: testing 
time for theories}, Nature 398, 199 (1999).

\bibitem{nggwi} Ng, Y.J.~\& van Dam H., {\it Measuring the 
foaminess of space-time with gravity-wave interferometers},
Found.~Phys.~30, 795-805 (2000).

\bibitem{polonpap} Amelino-Camelia G., {\it Are we at 
the dawn of quantum-gravity phenomenology?},
Lect.~Notes~Phys.~541, 1-49 (2000).

\bibitem{string1} Green M.B.,
Schwarz, J.H.~\& Witten E., {\it Superstring theory}
(Cambridge Univ.~Press, Cambridge, 1987).

\bibitem{string2} Polchinski J., {\it String theory}
(Cambridge Univ.~Press, Cambridge, 1998).

\bibitem{cqgab} Ashtekar A., 
{\it Quantum mechanics of geometry}, gr-qc/9901023.

\bibitem{cqgcar} Gaul M.~and Rovelli C.,
{\it Loop Quantum Gravity and the Meaning of Diffeomorphism Invariance},
Lect.~Notes~Phys.~541, 277-324 (2000).

\bibitem{cqglee} Smolin L.,
{\it The new universe around the next corner},
Physics World 12, 79-84 (1999).

\bibitem{ehns} Ellis J., Hagelin J.S., Nanopoulos D.V.~and Srednicki M.,
{\it Search for violations of quantum mechanics},
Nucl.~Phys.~B241, 381-405 (1984).

\bibitem{elmn1} Kostelecky V.A.~and Potting R.,
{\it CPT, strings, and meson factories},
Phys.~Rev.~D51, 3923-3935 (1995).

\bibitem{elmn2} Huet P.~and Peskin M.E.,
{\it Violation of CPT and quantum mechanics
in the $K_0-\bar{K_0}$ system},
Nucl.~Phys.~B434, 3-38 (1995).

\bibitem{elmn3}
Ellis J., Lopez J., Mavromatos N.E., 
Nanopoulos D.~and CPLEAR Collaboration,
{\it Test of CPT Symmetry and Quantum Mechanics with
Experimental data from CPLEAR},
Phys.~Lett.~B364, 239-245 (1995).

\bibitem{grbgac} Amelino-Camelia G., 
Ellis J., Mavromatos N.E., 
Nanopoulos D.~and Sarkar S., 
{\it Tests of quantum gravity from observations 
of $\gamma$-ray bursts}, 
Nature {393}, 763-765 (1998).

\bibitem{saulson} Saulson P.R., {\it Fundamentals of interferometric 
gravitational wave detectors} (World Scientific, Singapore, 1994).

\bibitem{rwold} Radeka V.,
{\it Low-Noise Techniques in Detectors},
Ann.~Rev.~Nucl.~Part.~Sci.~38, 217-277 (1988).

\bibitem{garaythermal} Garay L.J., {\it Space-time foam 
as a quantum thermal bath},
Phys.~Rev.~Lett.~80, 2508-2511 (1998).

\bibitem{nautilus} Astone P.~{\it et al},
{\it Upper limit for a gravitational-wave
stochastic background with the EXPLORER
and NAUTILUS resonant detectors},
Phys.~Lett.~B385, 421-424 (1996).

\bibitem{micgwb} Maggiore M.,
{\it Gravitational
Wave Experiments and Early Universe Cosmology},
Physics Reports 331, 283-367 (2000).

\bibitem{ligo} Abramovici A.~{\it et al}, 
{\it LIGO: The Laser Interferometer Gravitational-Wave
Observatory},
Science {256}, 325-333 (1992).
(Updated information
on expected sensitivity of an advanced phase
of the LIGO interferometer can be found
at WWW site http://www.ligo.caltech.edu/~ligo2/.)
 
\bibitem{virgo} Caron  B.~{\it et al},
{\it The Virgo interferometer},
Class.~Quantum Grav.~14, 1461-1469 (1997).
(Details on the sensitivity objectives of VIRGO can be
found in the document VIR-NOT-PER-1390-51,
available at the WWW site http://www.virgo.infn.it/.)

\bibitem{lisa} Danzmann K.,
{\it LISA: Laser interferometer space antenna for 
gravitational wave measurements},
Class.~Quantum Grav.~13, A247-A250 (1996).

\bibitem{thooftnew} G.~'t Hooft, {\it Quantum gravity as a dissipative
deterministic system}, 
Class.~Quantum Grav.~16, 3263-3279 (1999).

\bibitem{longnat} Amelino-Camelia G., 
{\it Phenomenological description of space-time foam},
gr-qc/0104005.

\end{thebibliography}
\end{document}